\documentstyle[12pt,twoside,fleqn,QM95,psfig]{article}
   \newcommand{\lsim}{\buildrel < \over {_\sim}}
   \newcommand{\gsim}{\buildrel > \over {_\sim}}

\newcommand{\AmS}{{\protect\the\textfont2
  A\kern-.1667em\lower.5ex\hbox{M}\kern-.125emS}}

\title{ Approach to parton equilibration \thanks
{This work was supported by the Director, Office of Energy
Research, Division of Nuclear Physics of the Office of High Energy and Nuclear
Physics of the U.S. Department of Energy under Contract Nos. DE-AC03-76SF00098,
and by the U.S.-Hungarian Science and Technology Joint Fund J.F.No.378.
}}
\author{Xin-Nian Wang \\ \vspace{12pt}
Nuclear Science Division, MS70A-3307, Lawrence Berkeley Laboratory\\
University of California, Berkeley, CA 94720}
       
\begin{document}
\maketitle

\begin{abstract}

  Perturbative QCD-based models of parton production and equilibration
in ultrarelativistic heavy ion collisions are reviewed with an emphasis
on the treatment of quantum interference effects. Uncertainties in the
initial parton production and their effects on later parton
equilibration are considered. Probes of early parton dynamics
are also discussed.

\end{abstract}

\bigskip

\section{INTRODUCTION}

In the last few years, there have been enormous new interests in
a perturbative QCD-inspired description of ultrarelativistic heavy
ion collisions based on the parton model \cite{KGQM93}. The main 
argument for such a  treatment of high-energy heavy ion collisions 
lies in the the nuclear structure at different scales. When the 
transverse momentum transfer involved in each nucleon-nucleon 
collision is small, $p_T {\ \raisebox{2.75pt}{$<$}\hspace{-9.0pt}
\raisebox{-2.75pt}{$\sim$}\ }\Lambda_{\rm QCD}$, effective models
based on,{\em e.g.}, meson-exchange and resonance formation
are sufficient to describe multiple interactions between hadrons, 
in which parton structure of the hadrons cannot  yet be resolved.
Those coherent (with respect to partons inside a hadron) 
interactions lead to collective behaviors in low and intermediate 
energy heavy ion collisions as  first observed in Bevalac 
experiments \cite{BEVL} and recently in experiments at the AGS 
energy \cite{STACHEL}. However, when $p_T$ becomes large enough 
to resolve individual partons inside a nucleon, the dynamics is 
best described on the  parton level via perturbative QCD (pQCD). Though 
hard  parton interactions occur at CERN-SPS energies 
($\sqrt{s}\lsim$ 20 AGeV), they play a  negligible role in the global 
features of heavy ion collisions.  However, at collider energies 
($\sqrt{s} \gsim 100$ AGeV) the importance of hard or semihard 
parton  scatterings is clearly seen in high-energy $pp$ and 
$p\bar{p}$ collisions \cite{WANGPP}. They are therefore also 
expected to be dominant in heavy ion collisions at  RHIC and LHC 
energies \cite{JBAM,KLL}.

Hard or semihard interactions happen in a very short
time scale and they generally break color coherence inside an
individual nucleon. After the fast beam partons pass through
each other and leave the central region, a dense partonic 
system will be left behind which is not immediately in thermal 
and chemical equilibrium. Partons inside such a system 
will then further interact with each other and equilibration 
will eventually be established if the interaction is frequent
enough among the sufficiently large number of initially produced
partons. Due to the asymptotic behavior of QCD, production 
rates of hard and semihard partons are calculable via pQCD 
during the initial stage of heavy ion collisions. The 
color screening mechanism in the initially produced dense partonic
system may make it also possible to use pQCD to investigate
the thermal and chemical equilibration of the system. 
In this talk, I will give a brief review of parton production 
in high energy heavy ion collisions. Then I will concentrate on
recent developments in early parton dynamics. I will also discuss
the uncertainties in the initial condition of parton equilibration and 
their consequences in the formation of a quark-gluon plasma.

\section{INITIAL PARTON PRODUCTION}

In a parton model, a nucleus can be considered as an assemble of
interacting partons. Because of vacuum fluctuations, a valence quark
can radiate many off-shell gluons, quarks and antiquarks. These partons, 
if undisturbed, will reassemble back to the valence quark. However, a
hard scattering can break the coherence and bring these virtual partons
on shell.  The parton number density $f_{a/A}(x,Q^2)$ thus depends on 
the resolution of the probe or the momentum transfer $Q^2$ of the 
collision. Similarly, a hard scattering can also lead to final 
state radiations off the scattered partons which can be described 
by fragmentation functions $D_{a/b}(z,Q^2)$. In the leading logarithmic
approximation in an axial gauge, the interference between multiple
radiations and between initial and final state radiations can be
neglected. The amplitude for successive radiations then has a simple
ladder structure and both the parton distribution  functions 
$f_{a/A}(x,Q^2)$ and the fragmentation functions follow the 
Altarelli-Parisi evolution equations \cite{AP} with respect to $Q^2$.

Using the parton distribution functions $f_{a/A}(x,Q^2)$, 
parton scattering rates in nucleus-nucleus collisions can be
written as
\begin{equation}
        \frac{dN_{jet}(b)}{dp^2_Tdy_1dy_2}=K\int d^2{\bf r} \sum_{a,b}
        x_1 f_{a/A}(x_1,p_T^2,{\bf r}) x_2 f_{b/B}(x_2,p_T^2,{\bf b-r})
        \frac{d\sigma_{ab}}{d\hat{t}}, \label{eq:njet}
\end{equation}
where $d\sigma_{ab}$ is the cross section for parton-parton
scatterings, $y_1$ and $y_2$ are the rapidities of
the scattered partons, $x_1$ and $x_2$ are the light-cone
momentum fractions carried by the initial partons, and the
summation runs over all parton species. The factor $K\approx 2$ 
accounts for next-to-leading order effects. The parton distribution
density of a nucleus is,
\begin{equation}
        f_{a/A}(x,Q^2,{\bf r})=t_A({\bf r})S_{a/A}(x,{\bf r})f_{a/N}(x,Q^2),
                        \label{eq:fx}
\end{equation}
where $t_A({\bf r})$ is the thickness of the nucleus which is normalized
to $\int d^2{\bf r}t_A({\bf r})=A$, $f_{a/N}(x,Q^2)$ is the parton
structure function of a nucleon and $S_{a/A}(x,{\bf r})$ 
accounts for the nuclear modification of parton distributions. 

\begin{figure}
\centerline{\psfig{figure=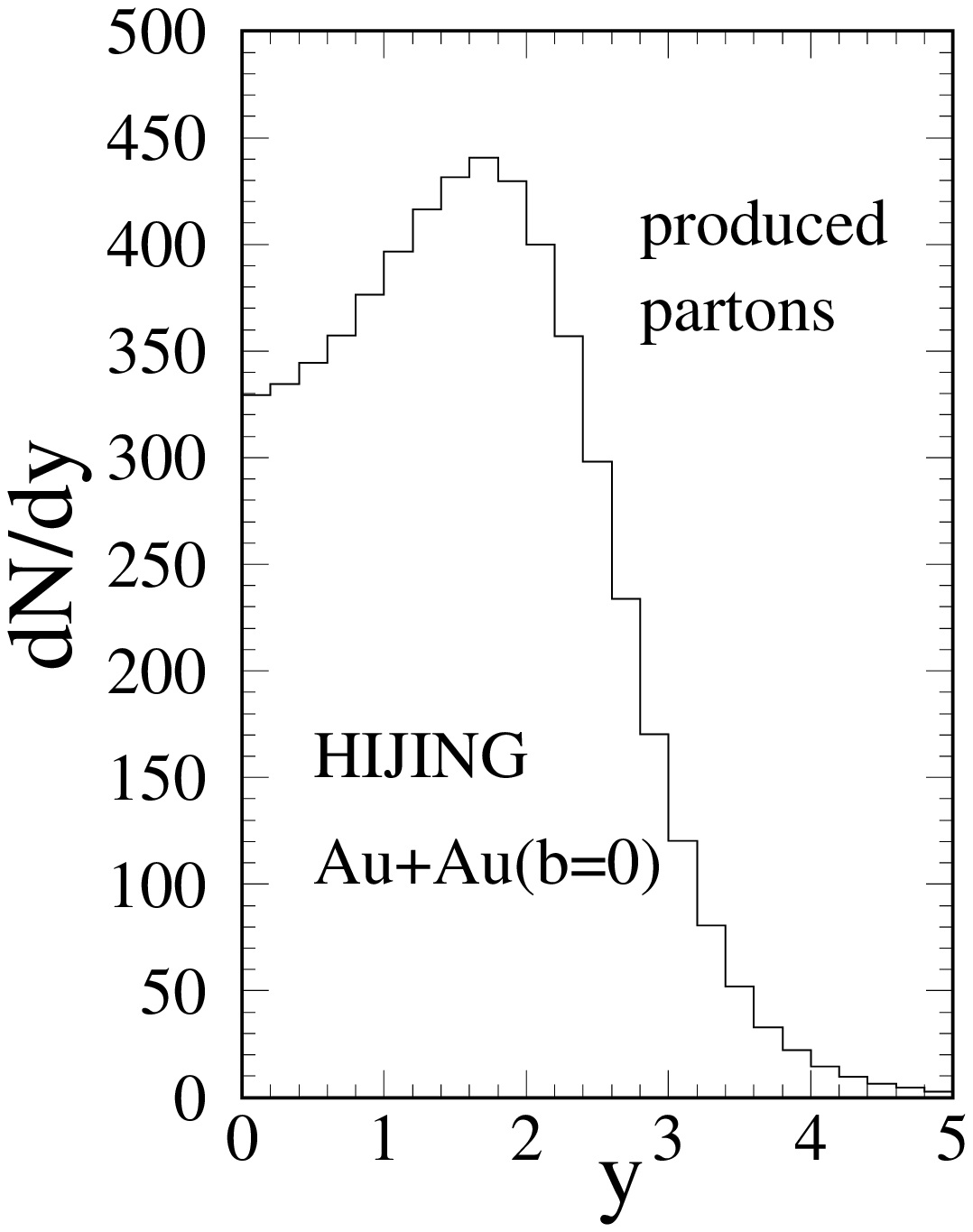,width=2.5in,height=3in}
\hspace{0.5in}\raisebox{-0.45in}{\psfig{figure=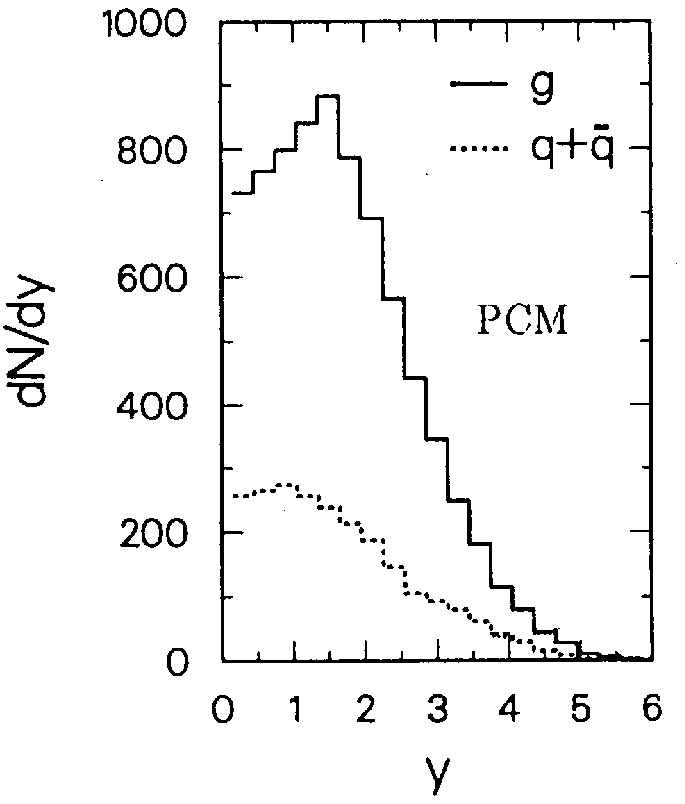,width=3in,height=3.4in}}}
\caption{Rapidity distributions of produced partons given by HIJING 
and PCM calculations for central $Au+Au$ collisions at RHIC energy}
\label{fig1}
\end{figure}

Many models have been developed to simulate parton production
based on Eq.~(\ref{eq:njet}) \cite{HIJING,KGBM,AMELIN,RANFT}. 
In particular, HIJING and PCM have applied the techniques of 
renormalization group improved perturbative QCD, initially 
developed to study jet formation in $e^+e^-$ and $p\bar{p}$ 
collisions \cite{WEBB,SJOS}, to simulate initial and final 
state radiations associated with hard or semihard parton 
scatterings. Shown in Fig.~\ref{fig1} are the rapidity densities
of produced partons given by HIJING and PCM \cite{KG92}. Both models predict
a large number of produced partons which will form a dense partonic
gas as the initial condition immediately after the overlap of the
two colliding nuclei with formation time, $1/p_T\sim 0.2$ fm/$c$. 
To ensure the credibility of these predictions, all models have 
been checked against the existing data of $pp$ and $p\bar{p}$ 
collisions. It has been demonstrated that many aspects of 
multi-particle production in $pp$ and $p\bar{p}$ collisions 
can be accounted for by the onset of semihard parton scatterings 
in high energy hadronic collisions.

Even though existing $pp$ and $p\bar{p}$ data can provide many 
constraints on these pQCD motivated phenomenological models, 
they still cannot uniquely determine all the parameters, giving 
rise to many uncertainties in the initial parton production:

(1)To regulate the infrared divergences in the QCD cross section
of minijet production, a $p_T$ cut-off $p_0$ has to be introduced. 
This cut-off and the corresponding soft inclusive cross section 
$\sigma_{\rm soft}$ for interactions with smaller $p_T<p_0$ 
are constrained by the energy dependence of the total $pp$ and $p\bar{p}$ 
cross sections.  These two parameters cannot be determined uniquely 
from the phenomenology of currently measured observables \cite{WANG91}.
To reduce the uncertainties, measurement of two-particle correlation 
functions in azimuthal angle in the transverse plane was 
proposed \cite{WANG92} to further constrain $p_0$ and 
$\sigma_{\rm soft}$.

(2)Since most of produced partons are gluons, their number
is sensitive to the gluon distribution function $f_g(x)$ at small $x$.
So far there is still no precise measurement of the gluon distribution
function in the small $x$ region where most of the minijets originate
at RHIC and LHC energies. Though there have been theoretical attempts 
to study gluon distribution in a nucleus \cite{EQW94,LMRV}, there is 
no direct experimental measurement of nuclear shadowing of the gluon 
distribution.  Future experiments at HERA might provide information 
on the gluon distribution. In addition systematical measurements of 
$pp$, $pA$ and $AA$ interactions at RHIC can be used to study 
gluon shadowing \cite{WG92}.

(3)The treatment of primary-secondary parton scatterings can also
lead to differences in the initial parton production. Because of
formation time effect which I will discuss later, most of the 
produced partons have no time to rescatter again with another
oncoming beam parton before they pass through. Proper inclusion
of such formation time effect will suppress primary-secondary
parton scatterings \cite{KEXW94a,WUHAN}. Systematic studies
of $pA$ collisions at RHIC will provide a quantitative estimate
of this formation time effect.

(4)Finally, parton production from soft processes is also very
model dependent. In parton cascade model, soft interactions are
treated as elastic parton scatterings with a regularized distribution
at small $p_T$. Such soft interactions with large cross 
section $\sigma_{\rm soft}$ can lead to many produced
partons, especially when the long interaction time is not taken
into account. In models with a string phenomenology, it is argued
that soft interactions, though not calculable in pQCD, produce
coherent gluons which form strings between leading partons. These
strings can materialize into soft partons but at rather longer
time scale of more than 1 fm/$c$. Color screening by the hard
partons produced earlier can also reduce the soft parton production
from the coherent color field \cite{KEMG93}.

\section{EARLY PARTON DYNAMICS}

The formation time of parton production can be estimated
via uncertainty principle. Thus model calculations can also
give the space-time history of parton production. Not surprisingly,
it is found \cite{KEXW94a} that most of partons are produced
within 0.4 fm/$c$ after the complete overlap of two colliding
nuclei. The parton system will then undergo further interactions 
and free-streaming. Neglecting parton rescatterings in this period 
of time, the kinematic separation of partons with different 
rapidities in a cell establishes local momentum isotropy at the 
time of the order of $\tau_{\rm iso}$ \cite{KEXW94a,BDMTW}.
Given this initial time $\tau_{\rm iso}$, the parton rapidity 
density $dN/dy$ and the average transverse momentum $\langle k_T\rangle$, 
the initial parton number and energy density can be estimated via 
Bjorken formula,
\begin{equation}
n_0 = \frac{1}{\pi R^2_{A} \tau_{\rm iso}} \frac{dN}{dy}\; ,
\quad \varepsilon_0 = n_0 \frac{4}{\pi}\langle k_T\rangle. \label{eq:bj}
\end{equation}
Assuming further that the phase distributions have a factorized 
form, $f_i(k)=\lambda_i f_{\rm eq}(k,T)$, one can then estimate
the effective initial temperature and fugacities. Listed in Table 1 
are the initial conditions for central $Au+Au$ collisions at RHIC 
and LHC collider energies from HIJING calculations. 

\begin{table}
\begin{center}
\caption{Values of the relevant parameters characterizing the parton
gas at time $\tau_{\rm iso}$.}
\label{table1}
\begin{tabular}{lrr} 
&\makebox[1in]{$\;\;\;$RHIC} &\makebox[1in]{$\;\;$LHC} \hfill \\ \hline
$\tau_{\rm iso}$ (fm/$c$) &0.7\phantom{000}   &0.5\phantom{000} \hfill \\
$\varepsilon_0$ (GeV/fm$^3$)  &3.2\phantom{000}   &40\phantom{.0000} \hfill \\
$n_0$ (fm$^{-3}$)             &2.15\phantom{00}   &18\phantom{.0000}  \hfill \\
$\langle k_{\perp}\rangle$ (GeV)    &1.17\phantom{00}  &1.76\phantom{00} \hfill\\
$T_0$ (GeV)                   &0.55\phantom{00}  &0.82\phantom{00} \hfill\\
$\lambda_g^0$                 & 0.05\phantom{00} &0.124\phantom{0} \hfill\\ 
$\lambda_q^0$                 &0.008\phantom{0}  &0.02\phantom{00} \hfill \\
\hline
\end{tabular}
\end{center}
\end{table}
One can observe that the initial parton gas is rather hot reflecting 
the large average transverse momentum. However, the parton gas is still
undersaturated as compared to the ideal gas at the same temperature. The
gas is also  dominated by gluons with its quark content far below
the chemical equilibrium value, confirming the hot glue 
scenario \cite{SHUR}.

How this parton gas evolves toward equilibrium is under intense
investigation in the last few years. Tremendous progresses have
been made with different approaches. Parton cascade model
looks particularly interesting since it can provide a space-time
picture of the evolution toward a thermalized quark-gluon plasma.
Because of the classical nature of the cascade model, however,
it is rather difficult to simulate many subtle quantum interference 
effects which are  expected to be very important. For example, 
M. Gyulassy and I have studied the criteria for a semiclassical 
treatment of multiple parton scatterings. We found \cite{GWLPM1} 
that the billiard ball picture of multiple scatterings in a cascade 
model is valid only when the momentum transfer of each scattering is 
large enough to resolve the spatial separation of nearby scatterers 
which may not be true in the initial scatterings of the beam partons. 
Another important problem is the color interference in QCD and how it 
can be incorporated into semiclassical simulations.

\begin{figure}
\centerline{\psfig{figure=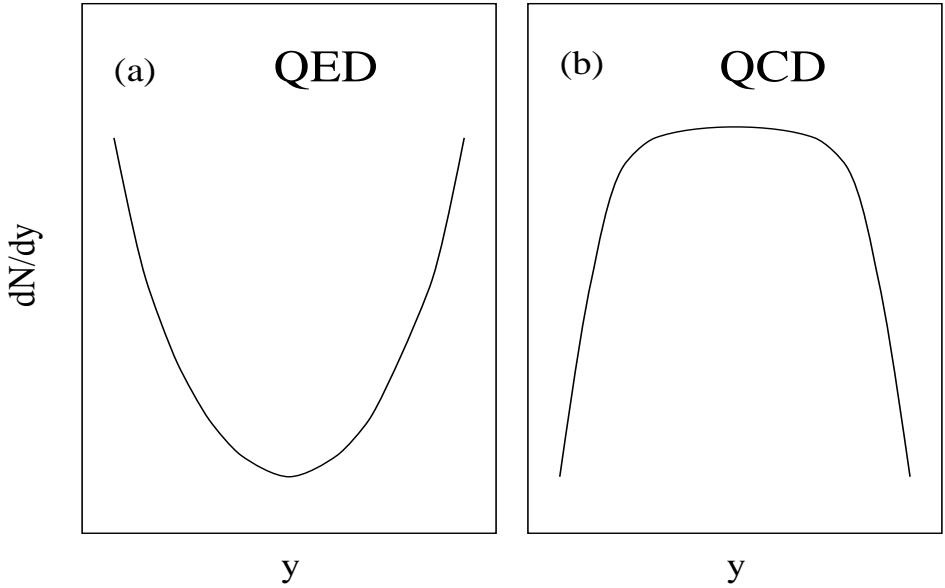,width=5in,height=3in}}
\caption{Illustrations of radiation spectrum in (a) QED and (b) QCD
  induced by a single scattering.}
\label{fig2}
\end{figure}

The importance of color interference can be best demonstrated
by comparing induced radiation in QED and QCD by a single scattering.
In QED case, the radiation amplitude can be separated into
initial state and final state radiation in the soft 
radiation limit,
\begin{equation}
  {\cal R}_1=i e\left ( \frac{\epsilon\cdot p_i}{k\cdot p_i}
  -\frac{\epsilon\cdot p_f}{k\cdot p_f}\right ),
\end{equation}
where $k$ and $\epsilon$ are the momentum and polarization
of the radiated photon, respectively, $p_i$ and $p_f$ are
the initial and final momentum of the charged particle.
In the high energy limit, the two terms in ${\cal R}_1$
cancel to the order ${\cal O}(1/\sqrt{s})$ in the central
region due to the destructive interference between initial and 
final state radiations. This gives rise to a valley in the photon's
rapidity distribution as illustrated in Fig.~\ref{fig2}(a). 
However, the cancellation will not be complete in QCD because 
gluons from initial and final state radiations have different 
colors due to the color exchange in the scattering. Furthermore, 
a gluon can also be emitted from the gluon propagator via the 
three gluon vertex. The resultant gluon spectrum \cite{JGGB}, 
\begin{equation}
  \frac{dn_g}{dyd^2k_\perp}=\frac{C_A\alpha_s}{\pi^2}
  \frac{q^2_\perp}{k^2_\perp({\bf q}_\perp-{\bf k}_\perp)^2}. 
  \label{eq:spec1}
\end{equation}
has a plateau in the central rapidity region, as shown
in Fig.~\ref{fig2}(b). This is very different from the QED case.

Similarly, destructive interference between induced radiations by
neighboring scatterings is also important. The radiation amplitude
induced by multiple scatterings in QED, for example, is,
\begin{equation}
  {\cal R}_m=i e \sum_{i=1}^m e^{ik\cdot x_i} 
    \left ( \frac{\epsilon\cdot p_i}{k\cdot p_i} -
    \frac{\epsilon\cdot p_{i-1}}{k\cdot p_{i-1}} \right ).
\end{equation}
In the eikonal limit ( particle propagating along a straight line),
the phase difference between any two terms can be written as,
\begin{equation}
  k\cdot (x_i-x_j)=L_{ij}/\tau(k); \quad \tau(k)=
  \frac{1}{\omega (1-\cos\theta)} \simeq \frac{2\omega}{k_\perp^2},
\end{equation}
where $L_{ij}$ is the distance (time) between scattering $i$ and $j$,
and $\tau(k)$ is the effective formation time. In the Bethe-Heitler
limit $L_{ij}\gg \tau(k)$, the total intensity of induced radiation 
is simply additive in the number of scatterings. However, when
$L_{ij}\ll \tau(k)$, the final state radiation of a scattering 
completely cancels with the initial state radiation from the previous
one. This so-called Landau-Pomeranchuk-Migdal (LPM) effect can also be
understood in terms of uncertainty principle: A radiated photon
( or gluon in QCD) must travel at least one wavelength $1/k_\perp$
in the transverse direction in order to be separated from its
parent particle. The radiation will be suppressed if the parent
particle suffers another scattering within time (distance) $\tau(k)$,
leading to the LPM effect.

Analysis in QCD gives similar results, although one has to take
into account different color structures associated with different
scatterings. In Ref. \cite{GWLPM1}, M. Gyulassy and I considered
induced radiation by multiple scatterings in the soft radiation
limit and neglected the situation where a radiated gluon can
suffer further interactions. We obtained a suppressed gluon spectrum
with effective formation time,
\begin{equation}
  \tau_{\rm QCD}(k)= \frac{C_A}{2C_2}\tau(k).
\end{equation}
Baier {\em et al}. \cite{BAIER} recently argued that those diagrams
which were neglected in Ref. \cite{GWLPM1} might be important
especially when a radiated real gluon travels with the beam
particle for a long time and then is knocked off later. This
argument will have important consequences in the energy loss
of a fast parton propagating in a QCD medium \cite{GWLPM2}. 
However, I will use the above analysis in terms of effective 
formation time to discuss induced radiation in the context of 
parton equilibration.

\section{PARTON EQUILIBRATION RATE}

I would like to emphasize that the above interference effects
happen on the matrix elements level. In particular, the LPM
effect involves the interference between initial and final state
radiations from different scatterings. A detailed analysis 
\cite{GWLPM1,GWLPM2} can show that the interference actually 
happens between two amplitudes in which the beam parton has 
completely different virtualities, {\em i.e.}, time-like in the 
final state radiation of one scattering and space-like in the initial 
state radiation of the previous one. This imposes a great difficulty for
a proper treatment of the quantum interference in a parton cascade
model in which a parton must {\em always} remain time-like between
two scatterings \cite{KGBM}.  A possible remedy for this is to consider
both initial and final state radiation associated with each scattering
together and include the interference effect by using a modified radiation
spectrum. The effective spectrum should suppress soft gluon radiation
whose formation time is larger than the mean-free-path of parton
scatterings. To demonstrate this, let us consider parton equilibration
in the form of rate equations \cite{BDMTW}. The analysis of QCD LPM 
effect \cite{GWLPM1,GWLPM2} has been done for a fast parton traveling 
inside a parton plasma. I will apply the results to radiations off thermal 
partons whose average energy is about $T$, since we expect
the same physics to happen.

We consider only the dominant process $gg\to ggg$ in the leading order. 
In order to permit the approach to chemical equilibrium, the reverse 
process, {\em i.e.}, gluon absorption, has to be included as well, 
which is easily achieved making  use of detailed balance. 
Radiative processes involving quarks have substantially smaller 
cross sections in pQCD, and quarks are considerably less 
abundant than gluons in the initial phase of the chemical evolution of 
the parton gas.  Here we are interested in understanding the basic 
mechanism underlying the formation of a chemically equilibrated 
quark-gluon plasma, and the essential time-scale.  We hence restrict 
our considerations to the dominant reaction mechanism for the 
equilibration of each parton flavor.  These are just four 
processes:
\begin{equation}
gg \leftrightarrow ggg, \quad gg\leftrightarrow
q\overline{q}.\label{eq:eq4}
\end{equation}
Other scattering processes ensure the maintenance of thermal
equilibrium $(gg\leftrightarrow gg, \; gq \leftrightarrow gq$, etc.)
or yield corrections to the dominant reaction rates 
$(gq\leftrightarrow qgg$, etc.). 

Restricting to the above reactions and assuming that elastic parton 
scatterings are sufficiently rapid to maintain local thermal equilibrium, 
the evolution of the parton densities  is governed by the master 
equations \cite{BDMTW}:
\begin{eqnarray}
\partial_{\mu}(n_gu^{\mu}) &= &
 \frac{1}{2}\sigma_3 n_g^2 \left( 1-\frac{n_g}{\tilde n_g}\right)
 -\frac{1}{2}\sigma_2 n_g^2 \left( 1 - \frac{n_q^2 \tilde n_g^2}
 {\tilde n_q^2 n_g^2}\right), \label{eq:eq5}\\
 \partial_{\mu} (n_qu^{\mu}) &= & \frac{1}{2}\sigma_2 n_g^2 
 \left( 1 - \frac{n_q^2 \tilde n_g^2}
 {\tilde n_q^2 n_g^2}\right), \label{eq:eq6}
\end{eqnarray}
where ${\tilde n_i}\equiv n_i/\lambda_i$ denote the densities 
with unit fugacities, $\lambda_i=1$, $\sigma_3$ and $\sigma_2$ 
are thermally averaged, velocity weighted cross sections,
\begin{equation}
\sigma_3 = \langle\sigma(gg\to ggg)v\rangle, \quad \sigma_2 =
\langle \sigma (gg\to q\bar q)v\rangle. \label{eq:eq7}
\end{equation}
We have also assumed detailed balance and a baryon symmetric
matter, $n_q=n_{\bar q}$. If we neglect effects of viscosity 
due to elastic scattering \cite{VISC}, we then have the 
hydrodynamic equation
\begin{equation}
{d\varepsilon\over d\tau} + {\varepsilon+P\over\tau} = 0. \label{eq:eq9}
\end{equation}

We further assume the ultrarelativistic equation of motion, 
$\varepsilon=3 P$. We can then solve the hydrodynamic equation,
\begin{equation}
  [\lambda_g + \frac{b_2}{a_2}\lambda_q]^{3/4} T^3\tau = \hbox{const.} \;\; , 
  \label{eq:eq10}
\end{equation}
and rewrite the rate equation in terms of time evolution of the
parameters $T(\tau)$, $\lambda_g(\tau)$ and $\lambda_q(\tau)$,
\begin{eqnarray}
\frac{\dot\lambda_g}{\lambda_g} + 3\frac{\dot T}{T} + \frac{1}{\tau} &=
&R_3 (1-\lambda_g)-2R_2 \left(1- \frac{\lambda_q^2}{\lambda_g^2} \right)
        \label{eq:eq11} \\
\frac{\dot\lambda_q}{\lambda_q} + 3\frac{\dot T}{T} + \frac{1}{\tau} &=
&R_2 {a_1\over b_1} \left( \frac{\lambda_g}{\lambda_q} -
\frac{\lambda_q}{\lambda_g}\right), \label{eq:eq12}
\end{eqnarray}
where the density weighted reaction rates $R_3$ and $R_2$ are defined as
\begin{equation}
R_3 = \textstyle{{1\over 2}} \sigma_3 n_g, \quad
R_2 = \textstyle{{1\over 2}} \sigma_2 n_g.  \label{eq:eq13}
\end{equation}
Notice that for a fully equilibrated system ($\lambda_g=\lambda_q=1$),
Eq. (\ref{eq:eq10}) corresponds to the Bjorken solution,
$T(\tau)/T_0=(\tau_0/\tau)^{1/3}$.

To take into account of the LPM effect in the calculation of
the reaction rate $R_3$ for $gg\rightarrow ggg$, 
we simply impose the LPM suppression of
the gluon radiation whose effective formation time $\tau_{\rm QCD}$
is much longer than the mean-free-path $\lambda_f$ of multiple 
scatterings to each $gg\rightarrow ggg$ process. 
In the mean time, the LPM effect also regularizes
the infrared divergency associated with QCD radiation. However,
$\sigma_3$ still contains infrared singularities in the gluon
propagators. For an equilibrium system one can in principle apply
the resummation technique developed by Braaten and Pisarski \cite{BP90}
to regularize the electric part of the propagators, though the
magnetic sector still has to be determined by an unknown magnetic
screening mass. Since we are dealing with a nonequilibrium system,
Braaten and Pisarski's resummation may not be well defined. As an
approximation, we will use the Debye screening mass \cite{BMW92},
\begin{equation}
\mu_D^2 = {6g^2\over \pi^2} \int_0^{\infty} kf(k) dk
=4\pi\alpha_s T^2\lambda_g, \label{eq:eq14}
\end{equation}
to regularize all singularities in the radiative cross section.

To further simplify the calculation we approximate the LPM 
suppression factor in Ref.~\cite{GWLPM1,GWLPM2} by a $\theta$-function,
$\theta(\lambda_f-\tau_{\rm QCD})$. The modified differential cross 
section for $gg\rightarrow ggg$ is then,
\begin{equation}
  \frac{d\sigma_3}{dq_{\perp}^2 dy d^2k_{\perp}}
  =\frac{d\sigma_{\rm el}^{gg}}{dq_{\perp}^2}\frac{dn_g}{dy d^2k_{\perp}}
  \theta(\lambda_f-\tau_{QCD})\theta(\sqrt{s}-k_{\perp}\cosh y),
\end{equation}
where the second step-function accounts for energy conservation, and 
$s=18T^2$ is the average squared center-of-mass energy of two 
gluons in a thermal gas. The regularized gluon density distribution
induced by a single scattering is,
\begin{equation}
  \frac{dn_g}{dy d^2k_{\perp}} =\frac{C_A\alpha_s}{\pi^2}
  \frac{q_{\perp}^2}{k_{\perp}^2[({\bf k}_{\perp}
    -{\bf q}_{\perp})^2 +\mu_D^2]}. \label{eq:dng}
\end{equation}
Similarly, the regularized small angle $gg$ scattering cross is,
\begin{equation}
  \frac{d\sigma_{\rm el}^{gg}}{dq_{\perp}^2}
  =\frac{9}{4}\frac{2\pi\alpha_s^2}{(q_{\perp}^2+\mu_D^2)^2}.
\end{equation}
The mean-free-path for elastic scatterings is then,
\begin{equation}
  \lambda_f^{-1}\equiv n_g\int_0^{s/4}dq_{\perp}^2
  \frac{d\sigma_{\rm el}^{gg}}{dq_{\perp}^2}
  =\frac{9}{8}a_1\alpha_s T\frac{1}{1+8\pi\alpha_s\lambda_g/9},
\end{equation}
which depends very weekly on the gluon fugacity $\lambda_g$. 

\begin{figure}
\parbox[b]{1.5in}{\caption{The scaled gluon production rate $R_3/T$ 
    (solid line) for $gg\rightarrow ggg$ and the quark production rate 
         $R_2/T$ (dashed line) for $gg\rightarrow q\bar{q}$
         as functions of the gluon fugacity $\lambda_g$ for 
         $\alpha_s$ = 0.3.}\label{fig3}\vspace{0.6in}} \hspace{0.4in}
     \psfig{figure=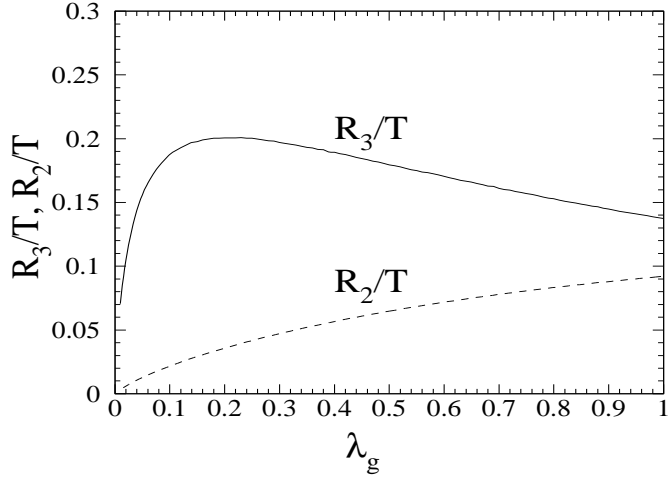,width=3.5in,height=2.5in}
\end{figure}

We can evaluate the integration numerically and find out
the dependence of $R_3/T$ on the gluon fugacity $\lambda_g$
as shown in Fig.~\ref{fig3}, for a coupling constant $\alpha_s=0.3$. 
The gluon production rate increases with $\lambda_g$ and then 
saturates when the system is in equilibrium. Overall, the gluon
equilibration rate is suppressed due to the inclusion of the LPM
effect on induced radiation.

The calculation of the quark equilibration rate $R_2$ for
$gg\rightarrow q\bar{q}$ is more straightforward. 
Estimate in Ref.~\cite{BDMTW} gives,
\begin{equation}
R_2 = {1\over 2}\sigma_2 n_g \approx 0.24 N_f \;\alpha_s^2 \lambda_g T
\ln (5.5/\lambda_g). \label{56}
\end{equation}
The dashed line in Fig.~\ref{fig3} shows the normalized 
rate $R_2/T$ for $N_f=2.5$, taking into account the reduced 
phase space of strange quarks at moderate temperatures, as a
function of the gluon fugacity.

\section{EVOLUTION OF THE PARTON PLASMA}

\begin{figure}
\centerline{\psfig{figure=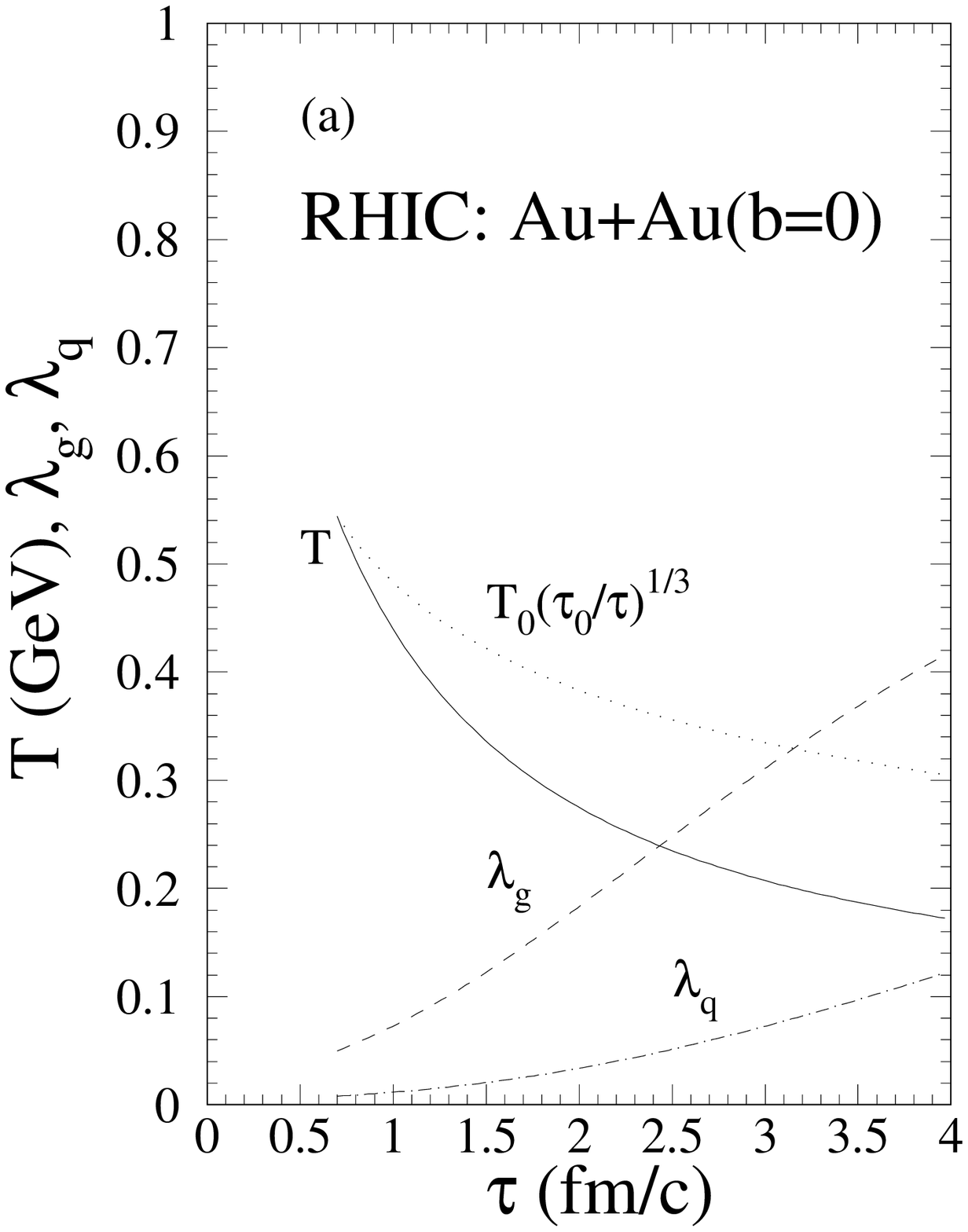,width=2.5in,height=3in}
  \hspace{0.5in}\psfig{figure=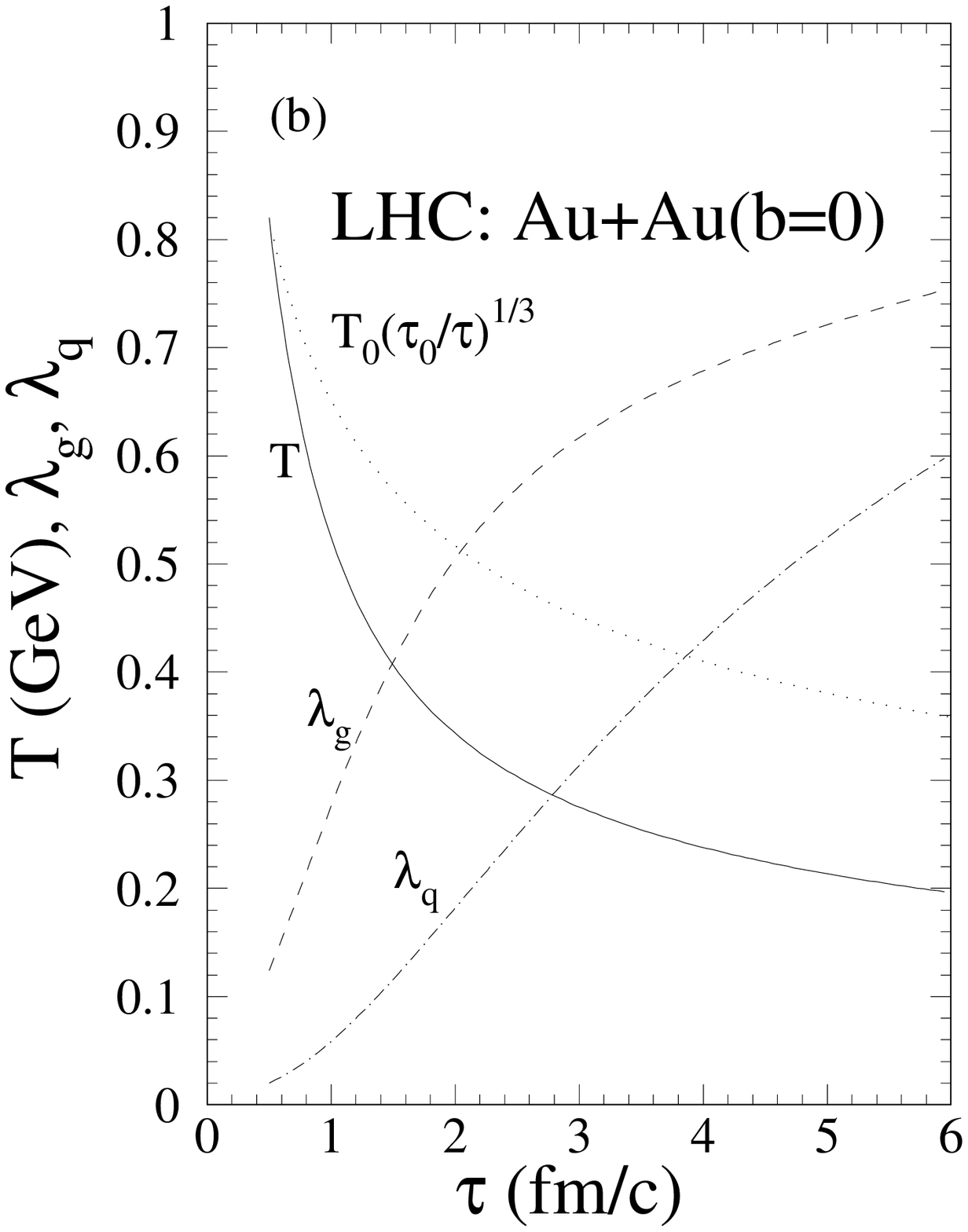,width=2.5in,height=3in}}
\caption{Time evolution of the temperature $T$ and the
         fugacities $\lambda_g$ and $\lambda_q$ of gluons and quarks in the
         parton plasma for central $Au + Au$ collisions at the (a) RHIC 
         and (b) LHC energies.
         The initial values for $T,\; \lambda_g$ and $\lambda_q$ are 
         determined from HIJING simulations and are listed in 
         Table~\protect\ref{table1}.}
\label{fig4}
\end{figure}

With the parton equilibration rates which in turn depend
on the gluon fugacity, we can solve the master equations 
self-consistently and obtain the time evolution of the temperature
and the fugacities. Shown in Fig.~\ref{fig4}
is the time dependence of $T$, $\lambda_g$, and $\lambda_q$
for initial conditions listed in Table~\ref{table1} at the RHIC and
LHC energies. We find that the parton gas cools considerably 
faster than predicted by Bjorken's scaling solution 
$(T^3\tau$ = const.) shown as dotted lines, because the production 
of additional partons approaching the chemical equilibrium 
state consumes an appreciable amount of energy. The 
accelerated cooling, in turn, slows down the chemical 
equilibration process, which is more apparent at the RHIC 
than at LHC energies. Therefore, the parton system can hardly
reach its equilibrium state before the effective temperature
drops below $T_c \approx 200$ MeV in a short period of time of
1-2 fm/$c$ at the RHIC energy. At the LHC energy, however, the parton
gas becomes very close to its equilibrium and the plasma
may exist in a deconfined phase for as long as 4-5 fm/$c$.
Another important observation is that quarks never
reach chemical equilibrium at both energies.
This is partially due to the small initial quark fugacity
and  partially due to the small quark equilibration rate.

In another recent development, E. Shuryak and L. Xiong \cite{XS93}
used the complete matrix elements of $gg\rightarrow gg+(n-1)g$
which include complete interference to calculate the parton 
equilibration rate. This is essentially a step beyond leading
logarithmic approximation used in most of the calculations. They found 
that inclusion of these higher order corrections not only
increases the initial parton production but also the
parton equilibration rate. However, it is expected that if
the formation time and virtual corrections (or unitarity) 
are included in this multiple gluon radiation processes,
the effect could become smaller.

\begin{figure}
\centerline{\psfig{figure=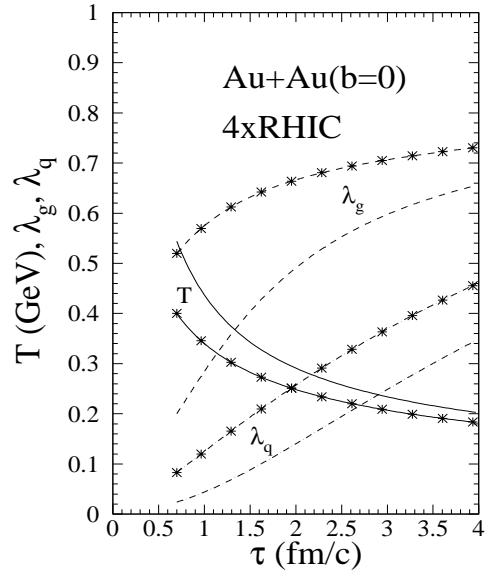,width=2.5in,height=3in}}
\caption{The same as in Fig.~\protect\ref{fig4}(a), except that the 
         initial parton densities are 4 times higher than given in 
         Table~\protect\ref{table1} with the same (ordinary lines),
         or reduced initial temperature, $T_0=0.4$ GeV (lines with stars)}
\label{fig5}
\end{figure}

As I have pointed out, there are still a lot of uncertainties
in the initial parton production. 
We can estimate the effect of the uncertainties in the 
initial conditions on the parton gas evolution by multiplying
the initial energy and parton number densities at the RHIC energy
by a factor of 4. This will result in the initial fugacities,
$\lambda_g^0=0.2$ and $\lambda_q^0=0.024$. With these high
initial densities, the parton plasma can evolve into a
nearly equilibrated gluon gas as shown in Fig.~\ref{fig5}.
The deconfined phase will also last longer for about 4 fm/$c$.
The system is still dominated by gluons with few quarks
and antiquarks as compared to a chemically equilibrated
system. If the uncertainties in the initial conditions
are caused by soft parton production from the color
mean field, the initial effective temperature will decrease.
Therefore, we can alternatively increase the initial 
parton density by a factor of 4 and decrease $T_0$ to
0.4 GeV at the same time. This leads to higher initial fugacities,
$\lambda_g^0=0.52$ and $\lambda_q^0=0.083$. As shown in
Fig.~\ref{fig5} by the curves with stars, this system evolves
faster toward equilibrium, however, with shorter life-time
in the deconfined phase due to the reduced initial temperature.

We thus can conclude that perturbative parton
production and scatterings are very likely to produce
a quark-gluon plasma ( or more specifically a gluon plasma)
in ultrarelativistic heavy ion collisions at the RHIC and LHC energies.
However, uncertainties in the initial conditions have to be
carefully examed in order to give a definite prediction of the 
quark-gluon plasma at the RHIC energy.

\section{PROBES OF EARLY PARTON DYNAMICS}

While some of the uncertainties may be reduced by theoretical
studies of the parton distribution in nuclei and initial
parton production, most of them have to be resolved experimentally.
There are many processes the measurements of which can provide 
experimental probes of the early parton dynamics. For example,
suppression of large $p_T$ jet or jet quenching 
\cite{WG92,MGMP90,GPTW,PGW95} due to the
final state interaction of large $p_T$ parton with the
parton gas can be used to study parton equilibration in
the early stage. Furthermore, the energy loss of the jets
and the alcoplanarity \cite{PC94,GUPTA} can also be used 
to study the thermal dynamical properties of the parton gas .

Among these hard probes, electromagnetical signals like thermal
photons and dileptons are considered more direct since they can
escape the dense matter without further interactions. They can thus
reveal the dynamics of initial parton production and equilibration.
In addition, open charm production can also be considered as
a direct probe since  charm quarks cannot be easily produced
during the mixed and hadronic phases of the dense matter
due to their heavy masses.  To leading order in pQCD, dilepton
production is dominated by $q\bar{q}\rightarrow \ell^+\ell^-$,
direct photon by $q(\bar{q})g\rightarrow q(\bar{q})\gamma$
and open charm by $gg\rightarrow c\bar{c}$ processes. As
pointed out by Strickland \cite{STLD}, measurements of these
direct thermal signals can tell us the relative ratio of
quark and gluon number densities in the early stage of
parton equilibration.

To demonstrate the sensitivity of these direct probes to
the initial condition of the parton evolution, let us consider
open charm production as an example. Charm production can be 
divided into three different contributions in the history of 
the evolution of the parton system: (1) initial production during 
the overlapping period ; (2) pre-thermal production from 
secondary parton scatterings during the thermalization, 
$\tau<\tau_{\rm iso}$; (3) and thermal production during the 
parton equilibration, $\tau>\tau_{\rm iso}$, in the expanding system.
The initial charm production can be calculated similarly to 
minijet production [cf. Eq.~(\ref{eq:njet})]. For thermal 
and pre-thermal production, the differential rate is\cite{BMXW92},
\begin{eqnarray}
  E\frac{d^3A}{d^3p}&=&\frac{1}{16(2\pi)^8}\int \frac{d^3k_1}{\omega_1}
  \frac{d^3k_2}{\omega_2}\frac{d^3p_2}{E_2}\delta^{(4)}(k_1+k_2-p-p_2)
  \nonumber \\
  & & \left[\frac{1}{2}g_G^2f_g(k_1)f_g(k_2)
  |\overline{\cal M}_{gg\rightarrow c\bar{c}}|^2+g_q^2f_q(k_1)f_{\bar{q}}(k_2)
  |\overline{\cal M}_{q\bar{q}\rightarrow c\bar{c}}|^2\right]
  , \label{eq:therm1}
\end{eqnarray}
given the phase-space density of the equilibrating partons, $f_i(k)$,
where $g_G$=16, $g_q=6N_f$, are the degeneracy factors for gluons and
quarks (antiquarks) respectively, 
$|\overline{\cal M}_{gg\rightarrow c\bar{c}}|^2$,
$|\overline{\cal M}_{q\bar{q}\rightarrow c\bar{c}}|^2$ are
the {\em averaged} matrix elements for $gg\rightarrow c\bar{c}$
and $q\bar{q}\rightarrow c\bar{c}$ processes. 
Due to small charm density, the Pauli blocking of the final charm 
quarks can be neglected. The corresponding charm spectrum is
\begin{equation}
  \frac{dN_c}{dyd^2p_{\perp}}=\pi R_A^2\int d\eta d\tau E\frac{d^3A}{d^3p}.
\end{equation}

\begin{figure}
\centerline{\psfig{figure=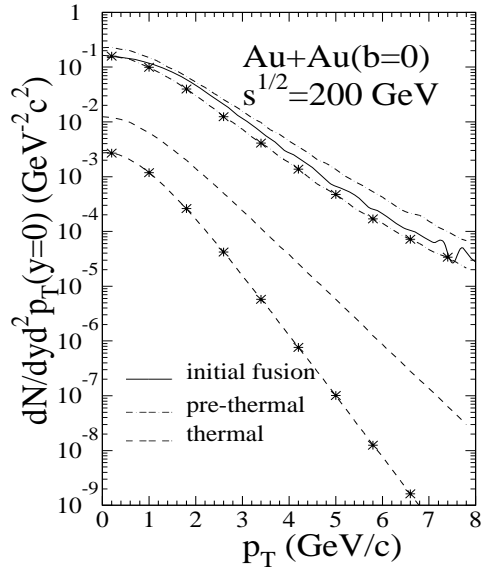,width=2.5in,height=3in}}
\vspace{-12pt}
\caption{The $p_{\perp}$ distribution of the initial (solid), pre-thermal
  (dot-dashed) and thermal (dashed) charm production for
  initial parton densities 4 times higher than HIJING estimate
  given in \protect\ref{table1} but with the same (ordinary lines), or
  reduced initial temperature, $T_0=0.4$ GeV (lines with stars).}
\label{fig6}
\end{figure}

For pre-thermal charm production, one can use the momentum
spectrum of initially produced partons and model the
phase distribution by introducing a correlation between 
momentum and space-time. This correlation was shown to
be important in the calculation of pre-thermal charm 
production \cite{LG94,LMW94}. For thermal production, one can
use the time dependence of temperature and fugacities as given
by the parton evolution equations. Shown in Fig.~\ref{fig6}
are the charm production rates as functions of $p_T$ with
the same initial conditions in Fig.~\ref{fig5}. Note that
with the increased initial parton density, open charm production
from pre-thermal and thermal stage is as important as the
initial production. Thermal production is more sensitive
to the variation of the initial temperature than the initial
fugacities. Therefore, by measuring charm enhancement, we
can probe the initial parton phase-space distribution,
initial temperature and equilibration time.

\section{CONCLUSIONS}

In this talk I reviewed the pQCD-based models of ultrarelativistic
heavy ion collisions. In this framework the reaction dynamics can
be described by perturbative parton scatterings and radiations.
Using model estimate of initial parton production, it is found
that there are enormous number of partons produced during
the overlap period of two colliding nuclei. Thereafter, local
isotropy in momentum distribution might be reached.  Parton
proliferation through induced radiation and  parton fusion
further drive the parton system toward a fully equilibrated
parton plasma. The quarks always lag behind gluons in reaching
their equilibrium. Due to the energy consumption by parton production, 
the parton gas cools down faster leading to a reduced plasma
life time of 4 - 6 fm/$c$.

Throughout the production and evolution of the partonic system,
interference effects play an important role in multiple collisions, 
the correct implementation of which will be a great
challenge to any {\em classical} cascade models.
Especially, the destructive interference among different amplitudes
of gluon radiation induced by multiple scatterings suppresses
soft gluons whose formation time is larger than the mean-free-path
of parton scatterings inside a QCD medium. A detailed analysis
of the radiation amplitudes reveals the underlying physics which
contradicts the intuitive picture of a classical cascade model.
One way to incorporate the LPM effect in the parton interaction 
simulations is to consider both the initial and final state 
radiation together for each scattering and impose the formation
time requirement, $\tau_{\rm QCD}<\lambda_f$, for the integration
over the phase-space of the radiated gluons. This will lead to
a reduced parton equilibration rate.

I also discussed the uncertainties in the initial conditions
and the experimental probes of the early parton evolution.
All in all, these uncertainties
arise from our ignorance of the nonperturbative physics
and our inability to calculate the soft processes in
the framework of QCD. In the pQCD-based models I have reviewed
here, the uncertainties really lie in the cut-off, $p_0$,
which supposes to separate nonperturbative soft interactions
from perturbative hard processes. Since soft and hard physics
do not have a definite boundary, the resultant parton production
from hard or semihard is very sensitive to the cut-off. The
accompanying soft parton production is not known in this
model and may only be estimated by simple models like the
color flux-tube model \cite{KEMG93}. There have been
recent developments in the field theory of particle production 
from mean-fields \cite{KLUGER,ELZE94}. The chaotic behavior of
nonabelian gauge fields may be intimately related to multiple parton 
production and fast gluon equilibration \cite{CHAOS}. Such a 
field theoretical approach to particle production could be the 
ultimate and consistent way to address the production and formation 
of a parton plasma in heavy ion collisions. 

In conclusion, parton equilibration is an exciting new subject
with many unsolved problems. I hope eventually, not far from now,
we can explicitly explain whether and how a quark-gluon plasma
is formed in a heavy ion collisions.

\noindent {\bf Acknowledgements}
I would like to thank my collaborators T.~S.~Bir\'o, K.~J.~Eskola,
M.~Gyulassy, P. L\'evai, B.~M\"uller and  M.~H.~Thoma who made 
significant contributions to the work I reviewed in this talk.
Stimulating discussions with M. Asakawa, K.~Geiger, H.~Heiselberg, 
J.~I.~Kapusta, 
L.~McLerran, E.~Shuryak, and  L.~Xiong are also
gratefully acknowledged.


\begin{thebibliography}{99}

\bibitem{KGQM93} K. Geiger, Nucl. Phys. {\bf A566}, 257c (1994),
        in Quark Matter'93, Proceedings of the 10th International
        Conference on Ultrarelativistic Nucleus-Nucleus Collisions,
        Borl\"ange, Sweden, June 20, 1993, edited by E. Stenlund,
        H.-\AA. Gustafsson, A. Oskarsson and I. Otterlund.
\bibitem{BEVL} H.~H~Gutbrod, A.~M.~Poskanzer and H.~G.~Ritter, Rep. Prog.
        Phys. {\bf 52}, 267 (1989).
\bibitem{STACHEL} J. Barrette {\em et al}., Phys. Rev. Lett. {\bf 73},
        2532 (1994).
\bibitem{WANGPP} X.-N.~Wang and M.~Gyulassy, Phys. Rev. D {\bf 45}, 
        844 (1992).
\bibitem{JBAM} J.~P.~Blaizot and A.~H.~Mueller, Nucl. Phys. 
         {\bf B289}, 847 (1987).
\bibitem{KLL} K. Kajantie, P. V. Landshoff, and Lindfors, Phys. Rev. Lett. 
        {\bf 59}, 2527 (1987); K.~J.~Eskola, K.~Kajantie and J.~Lindfors, 
        Nucl. Phys. {\bf B323}, 37 (1989).
\bibitem{AP}G.~Altarelli and G.~Parisi, Nucl. Phys. {\bf B126},
        298 (1977).
\bibitem{HIJING} X.-N. Wang and M.~Gyulassy, Phys. Rev. D {\bf 44},
        3501 (1991);  Comp. Phys. Commun. {\bf 83}, 307 (1994).
\bibitem{KGBM}K. Geiger and B. M\"{u}ller, Nucl. Phys. {\bf B369}, 
        600(1992); K. Geiger, Phys. Rev. D {\bf 47}, 133 (1993).
\bibitem{AMELIN} N.~S.~Amelin, E.~F.~Staubo, L.~P.~Csernai,
        Phys. Rev. D {\bf 46}, 4873 (1992). 
\bibitem{RANFT} I. Kawrakov, H.J. Mohring and J. Ranft, Phys. Rev. D {\bf 47},
        3849 (1993). 
\bibitem{WEBB} G.~Marchesini and B.~R.~Webber, Nucl. Phys. 
        {\bf B238}, 1 (1984).
\bibitem{SJOS} T. Sjostrand and M. van Zijl, Phys. Rev. D {\bf 36},
        2019 (1987).
\bibitem{KG92} K. Geiger and J. I. Kapusta, Phys. Rev. D {\bf 47}, 4905 (1992).
\bibitem{WANG91} X.-N.~Wang, Phys. Rev. D {\bf 43}, 104 (1991).
\bibitem{WANG92} X.-N.~Wang, Phys. Rev. D {\bf 46}, R1900 (1992),
        Phys. Rev. D {\bf 47}, 2754 (1993).
\bibitem{EQW94} K. J. Eskola, J. Qiu and X.-N. Wang, Phys. Rev. Lett.
        {\bf 72}, 36 (1994).
\bibitem{LMRV} L. McLerran and R. Venugopalan, Phys. Rev. D {\bf 49},
        3352 (1994).
\bibitem{WG92} X.-N. Wang and M.~Gyulassy, Phys. Rev. Lett. 
        {\bf 68}, 1480 (1992).
\bibitem{KEXW94a}K. J. Eskola and X.-N. Wang, Phys. Rev. D {\bf 49},
        1284 (1994).
\bibitem{WUHAN} X.-N. Wang, in proceedings of the Workshop on Finite
        Temperature QCD and Quark-gluon Transport Theory, Wuhan, China,
        April 18-26, 1994, edited by Lianshou Liu (World Scientific).
\bibitem{KEMG93}K. J. Eskola and M. Gyulassy, Phys. C {\bf 47}, 2329 (1993).
\bibitem{BDMTW} T.~S.~Bir\'o, E.~van~Doorn, B.~M\"uller, M.~H.~Thoma, 
        and X.-N.~Wang, Phys. Rev. C {\bf 48}, 1275 (1993).
\bibitem{SHUR} E. Shuryak, Phys. Rev. Lett. {\bf 68}, 3270 (1992).
\bibitem{GWLPM1} M.~Gyulassy and X.-N.~Wang, Nucl. Phys. {\bf B420},
        583 (1994).
\bibitem{JGGB} J.~F.~Gunion and G.~Bertsch, Phys. Rev. D {\bf 25},
        746 (1982).
\bibitem{BAIER} R. Baier, Yu. L. Dokshitzer, S. Peign\'e and D. Schiff,
        Report LPTHE-Orsay 94/98.
\bibitem{GWLPM2} X.-N.~Wang, M.~Gyulassy and M.~Pl\'umer, LBL-35980,
        Phys. Rev. D in press.
\bibitem{VISC} P. Danielewicz and M. Gyulassy, Phys. Rev. D {\bf 31},
        53 (1985); A. Hosoya and K. Kajantie, Nucl. Phys. {\bf B250},
        666 (1985); S. Gavin, Nucl. Phys. {\bf A435}, 826 (1985).
\bibitem{BP90}E. Braaten and R. D. Pisarski, Nucl. Phys. {\bf B337},
        569 (1990).
\bibitem{BMW92} T. S. Bir\'o, B. M\"uller, and X.-N. Wang, Phys. Lett. 
        {\bf B283}, 171 (1992).
\bibitem{XS93} L. Xiong and E. Shuryak, Phys. Rev. C {\bf 49}, 2207 (1994).
\bibitem{MGMP90} M.~Gyulassy and M.~Pl\"{u}mer, Phys. Lett. {\bf B243}, 
        432 (1990).
\bibitem{GPTW} M.~Gyulassy, M.~Pl\"umer, M.~H.~Thoma and X.-N.~Wang,
        Nucl. Phys. {\bf A538}, 37c (1992).
\bibitem{PGW95} M.~Pl\"umer, M.~Gyulassy, X.-N.~Wang, Proceedings
        of Quark Matter'95, Monterey, Jan. 9, 1995, edited by
         A. M. Poskanzer, J. W. Harris and L. S. Schroeder. 
\bibitem{PC94}J.-C. Pan, C. Gale, Phys. Rev. D {\bf 50}, 3235 (1994). 
\bibitem{GUPTA} S. Gupta, Report TIFR-TH-94-29, 1994.
\bibitem{STLD} M.~T.~Strickland, Phys. Lett. {\bf B331}, 245 (1994).
\bibitem{BMXW92} B. M\"uller and X. N. Wang, Phys. Rev. Lett. {\bf 68}, 
        2437 (1992).
\bibitem{LG94} Z. Lin and M. Gyulassy, Report CU-TP-638.
\bibitem{LMW94} P. L\'evai, B. M\"uller and X.-N. Wang, Report LBL-36594.
\bibitem{KLUGER}Y.~Kluger, J.~M.~Eisenberg, B.~Svetisky, F.~Cooper and
        E.~Mottola, Phys. Rev. Lett. {\bf 67}, 2427 (1991); Phys. Rev. 
        D {\bf 45}, 4659 (1992); {\em ibid.} D {\bf 48}, 190 (1993).
\bibitem{ELZE94}H.-T. Elze, CERN-TH-7131-93, hep-ph-9404215;
        CERN-TH-7297-94, hep-th-9406085.
\bibitem{CHAOS} B. M\"uller and A. Trayanov, Phys. Rev. Lett. {\bf 68},
        3387 (1992); C.~Gong,S.~G.~Matinian, B.~M\"uller and A.~Trayanov,
        Phys. Rev. D {\bf 49}, 607 (1994), 
        {\em ibid}. D {\bf 49}, 5629 (1994).
\end{thebibliography}
\end{document}